\documentclass[twocolumn,floatfix,times,trackchanges]{aastex63}

\def\peryr{\mbox{$\>\rm yr^{-1}$}}

\def\sun{\hbox{$\odot$}}
\def\earth{\hbox{$\oplus$}}
\def\lesssim{\mathrel{\hbox{\rlap{\hbox{%
 \lower4pt\hbox{$\sim$}}}\hbox{$<$}}}}
\def\gtrsim{\mathrel{\hbox{\rlap{\hbox{%
 \lower4pt\hbox{$\sim$}}}\hbox{$>$}}}}

\def\arcmin{\hbox{$^\prime$}}
\def\arcs{\hbox{$^{\prime\prime}$}}

\def\farcs{\hbox{$.\!\!^{\prime\prime}$}}

\def\micron{\hbox{$\mu$m}}

\newcommand{\REV}[1]{{#1}}

\usepackage{CJKutf8}
\usepackage{graphicx}
\usepackage{longtable} \LTcapwidth=\textwidth
\usepackage{multirow}
\usepackage{subfigure}
\usepackage{ulem}
\usepackage{soul}
\usepackage{lipsum, babel}
\usepackage{enumitem}
\usepackage{amsmath}
\usepackage{euscript}
\usepackage{verbatim}
\usepackage{booktabs} 

\setstcolor{red}

\accepted{\today}
\submitjournal{ApJL}

\shorttitle{Mass Assembly Turbulence Shock and Methanol}
\shortauthors{Hsu et al.}
\graphicspath{{./}{figures/}}

\begin{document}

\begin{CJK*}{UTF8}{bsmi}
\title{ALMASOP. Detection of Turbulence-induced Mass Assembly Shocks in Starless Cores}

\author[0000-0002-1369-1563]{Shih-Ying Hsu}
\email{seansyhsu@gmail.com}
\affiliation{Institute of Astronomy and Astrophysics, Academia Sinica, No.1, Sec. 4, Roosevelt Rd, Taipei 10617, Taiwan (R.O.C.)}

\author[0000-0012-3245-1234]{Sheng-Yuan Liu}
\email{syliu@asiaa.sinica.edu.tw}
\affiliation{Institute of Astronomy and Astrophysics, Academia Sinica, No.1, Sec. 4, Roosevelt Rd, Taipei 10617, Taiwan (R.O.C.)}

\author[0000-0001-8315-4248]{Xunchuan Liu}
\affiliation{Shanghai Astronomical Observatory, Chinese Academy of Sciences, Shanghai 200030, PR China}

\author[0000-0001-8077-7095]{Pak Shing Li}
\affiliation{Shanghai Astronomical Observatory, Chinese Academy of Sciences, Shanghai 200030, PR China}

\author[0000-0002-5286-2564]{Tie Liu}
\affiliation{Shanghai Astronomical Observatory, Chinese Academy of Sciences, Shanghai 200030, PR China}

\author[0000-0002-4393-3463]{Dipen Sahu}
\affiliation{Physical Research laboratory, Navrangpura, Ahmedabad, Gujarat 380009, India}

\author[0000-0002-8149-8546]{Kenichi Tatematsu}
\affiliation{Nobeyama Radio Observatory, National Astronomical Observatory of Japan, National Institutes of Natural Sciences, 462-2 Nobeyama, Minamimaki, Minamisaku, Nagano 384-1305, Japan}
\affiliation{Department of Astronomical Science, The Graduate University for Advanced Studies, SOKENDAI,
2-21-1 Osawa, Mitaka, Tokyo 181-8588, Japan}

\author[0000-0003-1275-5251]{Shanghuo Li}
\affiliation{School of Astronomy and Space Science, Nanjing University, 163 Xianlin Avenue, Nanjing 210023, People’s Republic of China }
\affiliation{Key Laboratory of Modern Astronomy and Astrophysics (Nanjing University), Ministry of Education, Nanjing 210023, People’s Republic of China}

\author[0000-0001-9304-7884]{Naomi Hirano}
\affiliation{Institute of Astronomy and Astrophysics, Academia Sinica, No.1, Sec. 4, Roosevelt Rd, Taipei 10617, Taiwan (R.O.C.)}

\author[0000-0002-3024-5864]{Chin-Fei Lee}
\affiliation{Institute of Astronomy and Astrophysics, Academia Sinica, No.1, Sec. 4, Roosevelt Rd, Taipei 10617, Taiwan (R.O.C.)}

\author[0000-0002-6868-4483]{Sheng-Jun Lin}
\affiliation{Institute of Astronomy and Astrophysics, Academia Sinica, No.1, Sec. 4, Roosevelt Rd, Taipei 10617, Taiwan (R.O.C.)}

\begin{abstract} 
Star formation is a series of mass assembly processes and starless cores, those cold and dense condensations in molecular clouds, play a pivotal role as initial seeds of stars.
With only a limited sample of known starless cores, however, the origin and growth of such stellar precursors had not been well characterized previously.
Meanwhile, the recent discovery of CH$_3$OH emission, which is generally associated with desorbed icy mantle in warm regions, particularly at the periphery of starless cores also remains puzzling. 
We present sensitive ALMA (Band~3) observations (at 3~mm) toward a sample of newly identified starless cores in the Orion Molecular Cloud.
The spatially resolved images distinctly indicate that the observed CH$_3$OH and N$_2$H$^+$ emission associated with these cores are morphologically anti-correlated and kinematically offset from each other.
We postulate that the CH$_3$OH emission highlights the desorption of icy mantle by shocks resulting from gas piling onto dense cores in the filaments traced by N$_2$H$^+$.
Our magnetohydrodynamic (MHD) simulations of star formation in turbulent clouds combined with radiative transfer calculations and imaging simulations successfully reproduced the observed signatures and reaffirmed the above scenario at work. 
Our result serves as an intriguing and exemplary illustration, a snapshot in time, of the dynamic star-forming processes in turbulent clouds.
The results offer compelling insights into the mechanisms governing the growth of starless cores and the presence of gas-phase complex organic molecules associated with these cores.
\end{abstract}


\section{Introduction}
\label{sec:Intro}
Starless cores are the precursors to stars; however, the processes governing their origin and growth remain poorly understood. 
Two main scenarios have been proposed to describe the evolution of starless cores: the ambipolar diffusion scenario and the supersonic turbulence scenario. 
The ambipolar diffusion model portrays the evolution of starless cores as a quasi-static gravitational accretion process \citep[e.g., ][]{1987_Shu}. 
In contrast, the supersonic turbulence model suggests that turbulence induces local compressions, leading to the formation of filaments and cores \citep[e.g., ][]{1999Maclow_driving,1999_Padoan,2001_Hartmann}). 
Turbulent fragmentation has been shown to satisfactorily explain the core population in clouds including Orion B, whereas clouds influenced by external pressure (e.g., Chamaeleon I) do not follow this model \citep[e.g., ][]{2024Fielder_G205M3}. 
While magnetohydrodynamic (MHD) turbulence is now widely believed to play a crucial role in starless cores \citep{2004Lazarian_review}, direct observational evidence linking turbulence with core evolution remains scarce.

Recent detections of gas-phase complex organic molecules (COMs) in starless cores introduce further puzzles. 
COMs, such as CH$_3$OH, are saturated organic molecules containing at least six atoms \citep{2009Herbst_COM_review}, primarily formed in the icy mantles of dust grains. 
The presence of gaseous COMs, CH$_3$OH in particular, would indicate desorption of icy mantles.
Thermal desorption mechanism, demonstrated by \citet{2023Hsu_ALMASOP} for example, successfully explained the intensity and extent of CH$_3$OH emission in the localized warm gas around a sample of protostellar cores in the Orion molecular cloud.
However, the desorption processes in starless cores remain debated, despite reports of gas-phase CH$_3$OH detection \citep[e.g., ][]{2020Scibelli_ColdCoresTaurus_COMs,2024Scibelli_Perseus_COM}. 

Several potential desorption mechanisms have been proposed, including cosmic ray ice sputtering \citep{2019Dartois_CR_sputtering,2021Wakelam_CR_sputtering} and reactive desorption \citep{2007Garrod_reactive-desorption,2013Vasyunin_reactive-desorption,2017Vasyunin_reactive-desorption,2018Chuang_reactive-desorption}.
Observationally, CH$_3$OH emission is found to be ubiquitous in chemical surveys toward the Taurus \citep{2020Scibelli_ColdCoresTaurus_COMs} and the Perseus clouds \citep{2024Scibelli_Perseus_COM}, suggesting the prevalence of desorption mechanisms.

In this Letter, we present our observations of CH$_3$OH and N$_2$H$^+$ of five dense starless cores in the Orion Molecular Cloud.
Section \ref{sec:methods} outlines our observational methods.
Based on the morphology and kinematics, we identify shock interfaces between dense filaments and the surrounding diffuse gas, as detailed in Section \ref{sec:results}. 
Our findings imply a framework in which turbulence-induced mass assembly shocks facilitate the release of methanol from ice surfaces to the gas phase. 
To support this framework, we incorporate core evolution simulations as is described in Section \ref{sec:ORION2}. 
Finally, in Section \ref{sec:Disc}, we discuss the broader implications of our results and their alignment with existing literature.


\section{Methods}
\label{sec:methods}

\begin{deluxetable*}{llllllllr}[tbph!]
\caption{\label{tab:coord} Observational parameters}
\tablehead{
\colhead{Source} & \colhead{Short}  & \colhead{Cloud} & \colhead{$\alpha_{2000}$} & \colhead{$\delta_{2000}$} & \colhead{BMAJ} & \colhead{BMIN} & \colhead{BPA} & \colhead{$v_\mathrm{LSR}$}
}
\startdata
G205.46-14.56M3 & G205M3 & B & 05h46m06.01s & -00d09m32.8s & 2\farcs{96} & 2\farcs{34} & 84\degr & 10.2 km~s$^{-1}$ \\
G208.68-19.20N2 & G208N2 & A & 05h35m21.04s & -05d00m57.9s & 2\farcs{93} & 2\farcs{28} & 87\degr & 11.0 km~s$^{-1}$ \\
G209.29-19.65S1 & G209S1 & A & 05h34m55.86s & -05d46m05.0s & 2\farcs{92} & 2\farcs{26} & 88\degr & 6.7 km~s$^{-1}$ \\
G209.94-19.52N & G209N & A & 05h36m11.32s & -06d10m44.8s & 2\farcs{92} & 2\farcs{26} & 88\degr & 8.0 km~s$^{-1}$ \\
G212.10-19.15N1 & G212N1 & A & 05h41m21.21s & -07d52m26.9s & 2\farcs{90} & 2\farcs{23} & 88\degr & 5.0 km~s$^{-1}$
\enddata
\tablecomments{
The coordinates are identified as the 3~mm continuum peak.
BMAJ, BMIN, BPA are the major axis, minor axis, and the position angle of the beam, respectively. 
The $v_\mathrm{LSR}$ is the system velocity, derived from the velocity shift of the N$_2$H$^{+}$ transition. 
}
\end{deluxetable*}

\subsection{Sample Selection}

To address these questions, we have selected a sample of five dense cores in the Orion Molecular Cloud, identified from the ALMA Survey of Orion PGCCs (ALMASOP) catalogue \citep{2020Dutta_ALMASOP}: G205.46-14.56M3, G208.68-19.20N2, G209.29-19.65S1, G209.94-19.52N, and G212.10-19.15N1 (hereafter referred to as G205M3, G208N2, G209S1, G209N, and G212N1, respectively).
The ALMASOP project targeted 72 extremely dense and compact 850 \micron\ continuum sources located at the Orion cloud, aiming to study the star formation at their early stages. 
Based on the observations, \citet{2020Dutta_ALMASOP} reported a catalogue of 23 starless cores and 56 Class 0/I protostellar cores. 
Our five targets represent the densest starless cores in the catalogue, identified through their significant detections \citep{2021Sahu_ALMASOP_presstellar}. 
The high densities imply that they are at the critical time for the subsequently forming central protostar. 
Details of the target cores are listed in Table~\ref{tab:coord}.

\subsection{Observations}

This study utilized data from the Atacama Large Millimeter/submillimeter Array (ALMA) Cycle~10 program (\#2023.1.01067.S, PI: Sheng-Yuan Liu) in Band~3 (3~mm). 
Observations were conducted with the 12-m array in the C-2 configuration, providing deprojected baseline lengths ranging from 5 to 104 k$\lambda$. 
The maximum recoverable scale in the observations was 28\farcs{8} and the final images achieved an angular resolution of $\sim$3\farcs{5}.

The receivers captured data across 16 spectral windows (SPWs), comprising 12 SPWs with bandwidths of 59 MHz and four SPWs with bandwidths of 117 MHz. 
The channel width is 61~kHz, corresponding to a velocity resolution of 0.2 km~s$^{-1}$. 
For this study, we focused on two SPWs having a bandwidth of 59~MHz centered at 93.164 GHz and 96.732 GHz, which covers the transitions of our targeted molecules, N$_2$H$^+$ and CH$_3$OH.
N$_2$H$^+$ is known to be a good tracer of the cold and dense gas since it is easily destroyed by gaseous CO and, therefore, can only survive in CO-depleted region \citep[e.g., ][]{2016Yamamoto_book}. 
In our spectral coverage, the multiple N$_2$H$^+$ $J=1–0$ hyperfine lines form three groups at $\sim$ 93.172 ($F_1=1-1$), 93.174 ($F_1=2-1$), and 93.176 ($F_1=0-1$) GHz. 
We used the one at 93.176 GHz ($F_1=0-1$) consisting of three transitions ($F=1-0$, $F=1-1$, and $F=1-2$) for the following two reasons. 
First, its hyperfine lines have frequencies that are indistinguishable within our spectral resolution, reducing the impact of line blending in the channel maps.
Second, this group had the weakest intrinsic line strength among the three; as a result, the emission was less optically thick. 
CH$_3$OH is the simplest COM and widely used as an indicator of complex organic chemistry. 
Three CH$_3$OH lines, namely the transitions $2_{0,2}–1_{0,1}~A$, $2_{1,2}–1_{1,1}~E$, $2_{0,2}–1_{0,1}~E$ having upper energies at 7, 13, and 20 K, respectively, were included in our spectral coverage.
We primarily used the one at 7~K as it has the strongest line intensity. 

Imaging was performed using the \texttt{tclean} task in CASA 6.5.4 \citep{casa:2022}. 
The mask parameter was set to ``auto-multithresh,'' the deconvolver was ``hogbom,'' and a Briggs robust weighting value of 2.0 was applied. 
The imaging thresholds were 0.2~mJy for the full-band ($\sim$0.857~GHz) continuum and 10~mJy for individual channels ($\sim$61~kHz per channel). 
The final images achieved a beam size $\sim$3\farcs{5} (1,400 au at 400~pc) and sensitivities of 0.025~mJy/beam for the full-band continuum and 2~mJy/beam for each channel.

\section{Results}
\label{sec:results}

\subsection{Morphology and Kinematics}

\begin{figure*}[htb!]
\centering
\includegraphics[width=.8\textwidth]{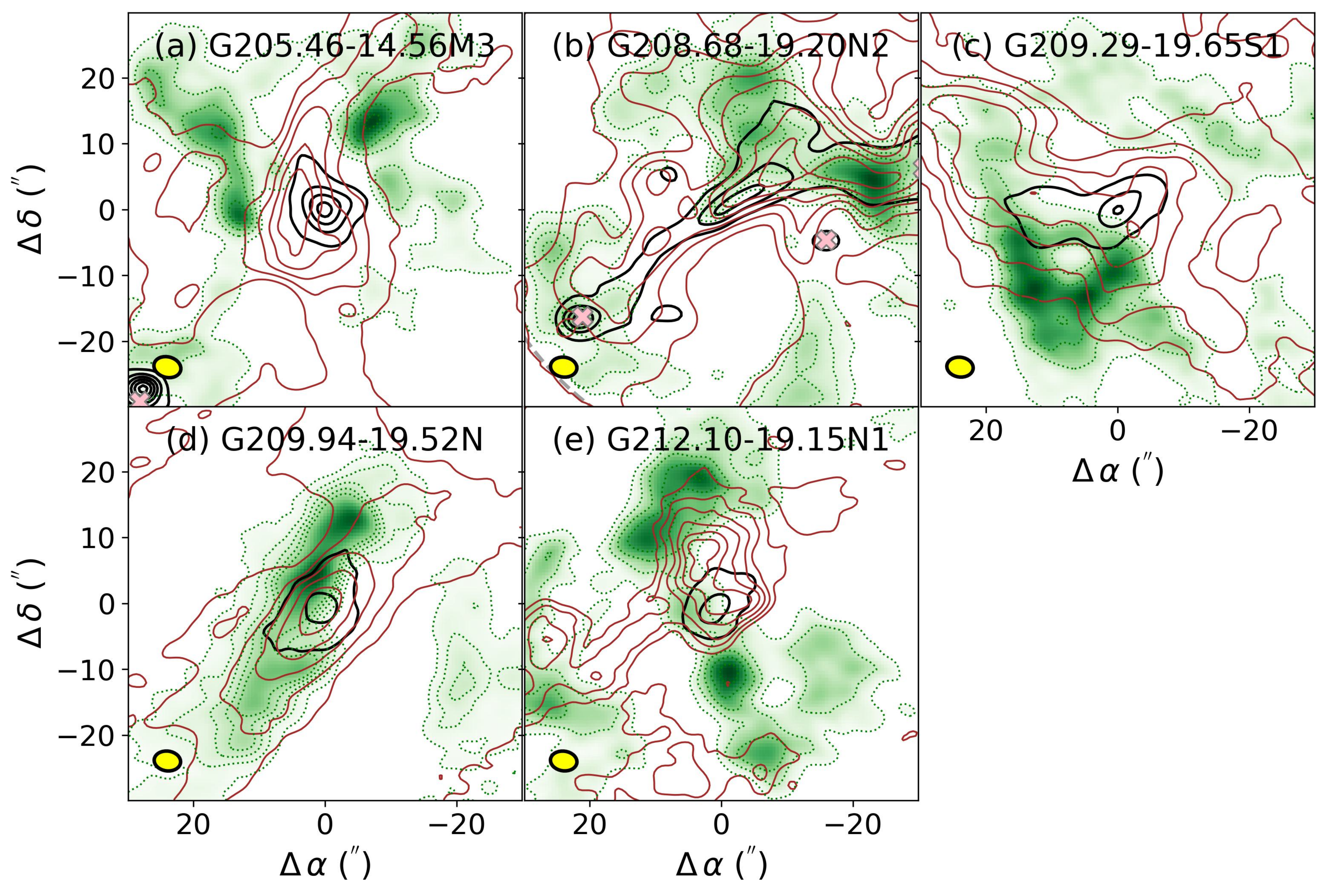}
\caption{
\label{fig:mom0_12m_Cycle~10}
3~mm continuum (black contours), integrated intensity of CH$_3$OH $E_\mathrm{u}=7$~K transition (green rasters and dotted contours), and integrated intensity of N$_2$H$^+$ $F_1=0-1$ transition (brown contours).
The contour levels for N$_2$H$^+$ (brown) are set as follows: [5, 45, 85, 125, 165] $\sigma$ for G208N2, [5, 17.5, 30, 42.5, 55] $\sigma$ for G209N, [5, 10, 15, 20, 25] $\sigma$ for G212N1, and [5, 15, 25, 35, 45] $\sigma$ for the remaining targets.
The contour levels for the 3~mm continuum (black) are set at [5, 15, 25, 35] $\sigma$. 
The contour levels for the CH$_3$OH (green dotted) are set at [5, 15, 25, 35, 45] $\sigma$.
The small ellipses in the bottom left corner represent the synthesized beam.
The pink crosses marked the nearby protostellar cores within the field, from east to west are G205.46-14.56N2 in panel (a) and HOPS 89 and HOPS 91  in panel (b). 
}
\end{figure*}

We imaged 3~mm continuum, N$_2$H$^+$ $J=1-0$ $F_1 = 0–1$, and CH$_3$OH $2_{0,2}$–$1_{0,1}~A^+$ with $E_\mathrm{u} = 7$~K line emissions in all five of our targets (Fig.\ref{fig:mom0_12m_Cycle~10}). 
In general, the N$_2$H$^+$ emission appears extended and encompasses the dust continuum, consistent with N$_2$H$^+$ tracing cold and dense gas.
However, the N$_2$H$^+$ and dust continuum peaks do not coincide, suggesting a possible depletion of N$_2$H$^+$ in the innermost and densest regions of these dense cold cores.
CH$_3$OH emission is distributed at the periphery of the N$_2$H$^+$ component and appears more fragmented.
This behavior is similar to that observed in the starless core L1544 on a scale of $\sim$2\arcmin~ \citep{2014Bizzocchi_L1544_CH2DOH, 2017Spezzano_L1544_ChemStructure}, where CH$_3$OH does not peak at the same positions where the dust continuum and the N$_2$H$^+$ emission peak.

\begin{figure*}[htb!]
\centering
\includegraphics[width=.8\textwidth]{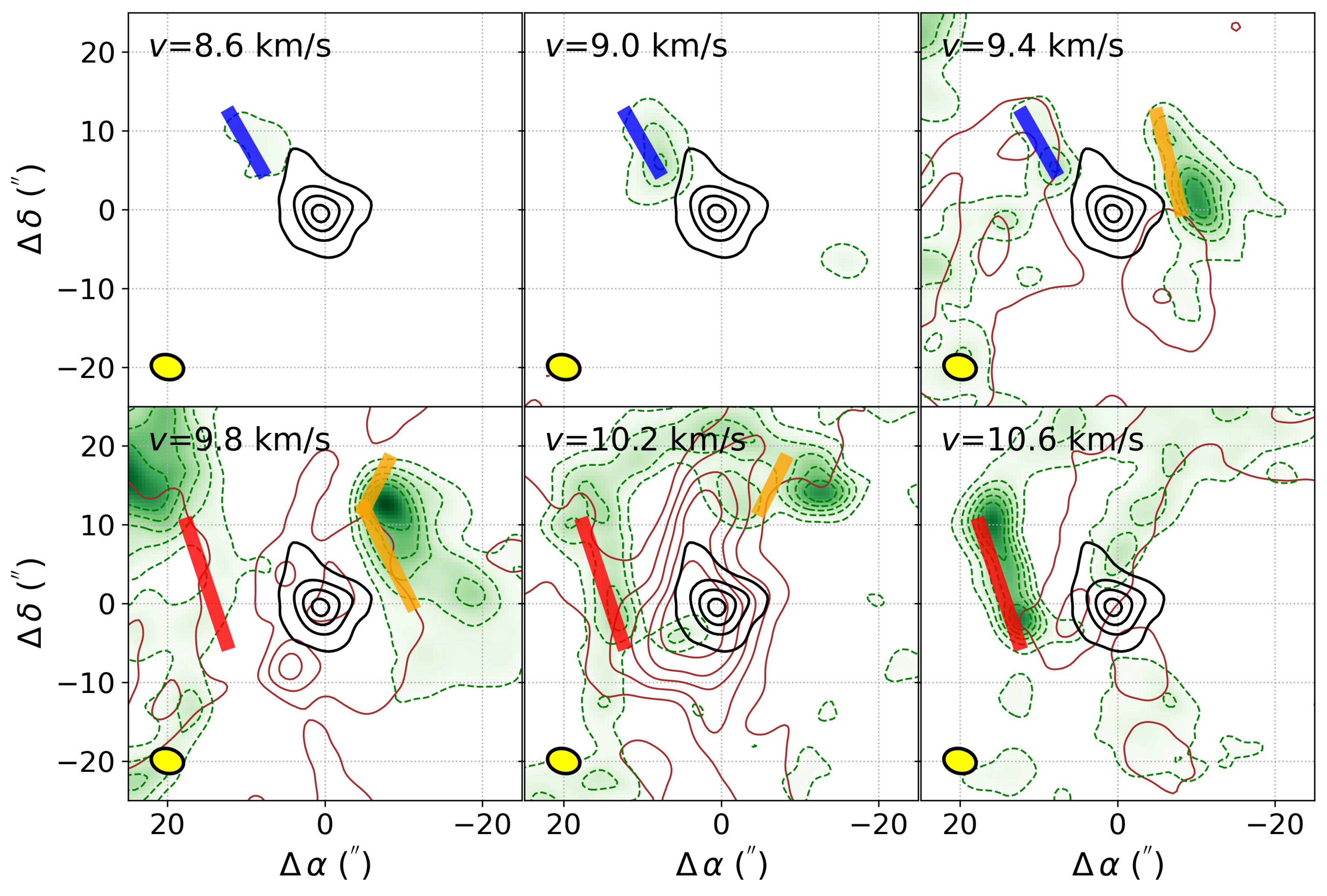}
\caption{
\label{fig:chn_shock_obs} 
Flux density channel map of G205M3. 
The green raster represents CH$_3$OH emission, the brown contours show N$_2$H$^+$ emission, and the black contours indicate the 3~mm continuum.
The contour levels for N$_2$H$^+$ (brown), CH$_3$OH (green), and 3~mm continuum (black) are set as  [5, 35, 65, 95, 125] $\sigma$, [5, 15, 25, 35, 45] $\sigma$, and [5, 15, 25, 35] $\sigma$. 
The three segments (blue, orange, and red) illustrate the interfaces between CH$_3$OH and N$_2$H$^+$ gas components having different velocity ranges. 
The origins of the coordinates are set at the 3~mm continuum peak position. 
The channel width of the observation is 0.2 km/s. In the figure we only presented the map for every two channels due to the limitation of the panel number. 
}
\end{figure*}

Interestingly, upon visually inspecting the channel map matrix, as exemplified in Fig. \ref{fig:matrix_G205M3}, we observe velocity differences between CH$_3$OH and its neighboring N$_2$H$^+$ counterpart.
For instance, Fig.~\ref{fig:chn_shock_obs} presents the channel maps of G205M3, where the velocity jumps are recognized more clearly.
Taking the interface marked by the red segment as an example, the CH$_3$OH component appears at $v=10.6$ km/s, while the N$_2$H$^+$ counterpart appears at $v=10.2$ km/s, inferring a velocity jump of 0.4 km/s. 
For the blue segment, the velocity jump reaches $9.4-8.6=0.8$ km/s. 
These velocity jumps of 0.4 -- 0.8 km/s are higher than the sound speed in cold cores (approximately 0.2 km/s at 10 K \citep{2007Bergin_review}), suggesting the existence of shocks at the interface between CH$_3$OH and N$_2$H$^+$ gas components \REV{within a field size of $\sim$50\arcs ($\sim$20,000~au)}. 
The N$_2$H$^+$ component, characterized by CO depletion, represents denser regions of the core.
The velocity jumps (0.4 -- 0.8 km/s) between gas components are consistent with the velocity dispersion (0.45 km/s) due to the turbulence activities at a cloud size of 20,000 au inferred from \citet{1981Larson_turbulence}. 

\subsection{Implications of the detected Shocks}

It is generally accepted that in cold starless cores CH$_3$OH is produced on dust grains and then liberated into the gas phase in the later protostellar stage \citep[e.g., ][]{2009Herbst_COM_review}), despite other pathways were also suggested \citep[e.g., reaction of CH$_3^+$ with water ice, ][]{2023Nakai_CH3OH_formation}.
Our findings suggest that the desorption leading to the observed CH$_3$OH gas is the result of the shocks induced by turbulence activities. 
Shocks could possibly cause the presence of COMs via thermal or mechanical processes.
In the thermal processes, a slow shock with a velocity jump of 0.4 km/s can heat dust grains from 10 to 30 K and trigger the desorption of grain mantles \citep{2001Dickens_TMC1,2015Soma_TMC1}. 
Alternatively, the ``mechanical removal'' of the icy mantle in oblique collisions between grains could lead to the presence of abundant gaseous COMs \citep{2022Kalvans_desorption}. 
We favor the latter interpretation, as the CH$_3$OH transition with higher upper energy ($E_\mathrm{u} = 20$ K) is either not detected or only weakly detected compared to the $E_\mathrm{u} = 7$ K transition (Figure \ref{fig:mom0_CH3OH_20K}).
Particularly, in the case of G208N2, our observed CH$_3$OH is distributed on the northeastern side, whereas \citet{2024Hirano_G208.68-19.20N2_dense} suggested that the southern edge of the filamentary cloud is relatively warmer, making thermal desorption of CH$_3$OH unlikely. 

We further postulate that the turbulence we observed drives the gas assembly onto dense cores within filaments. 
Previous studies have demonstrated that MHD turbulence can facilitate mass assembly in molecular clouds and lead to the formation of high-density cores \citep{2001Balsara_turbulent-MHD-model}. 
Observationally, a transition from a  turbulent parent cloud to the quiescent dense core was proposed to explain the coexistence of narrow and broad red-shifted NH$_3$ lines in the starless core TUKH122 \citep{2016Ohashi_TUKH122_turbulence}. 
Similarly, in two sources within the Cepheus-L1251 cloud, a velocity offset greater than 0.5 km/s between NH$_3$ and HC$_5$N was interpreted as evidence of a transition from the turbulent parental cloud to the quiescent dense region \citep{2017Keown_CepheusL1251}. 

We estimated the mass assembly rate ($\Dot{M}$) using our observations, following the relation:
\begin{equation}
\Dot{M} = \rho\ A\ \delta v, 
\end{equation}
where $\rho$ is the gas density, $A$ is the area, and $\delta v$ represents the assembly velocity. 
For each pixel, the area $A$ is assumed to be the physical size of the pixel, i.e., $A=$ (400 pc $\times$ 0\farcs{5})$^{2}$. 
We assumed that the assembly velocity is comparable to the velocity difference between the  quiescent dense core and its turbulent ambient, defined as $\delta v = v - v_\mathrm{LSR}$, where $v_\mathrm{LSR}$ is the systemic velocity of the dense core derived from the velocity shift of the N$_2$H$^+$ transition and $v$ corresponds to the frequency at which the CH$_3$OH line peaks. 
The velocity difference $\delta v$ was further multiplied by a factor of $\sqrt{3}$ to account for the projection. 
To estimate $\rho$, we first calculated the integrated brightness temperature ($W$) and derived the CH$_3$OH column density of the upper energy state ($N_\mathrm{u}$) using:
\begin{equation} 
N_\mathrm{u} = \frac{8\pi k_\mathrm{B}}{hc^3} \frac{\nu W}{A_\mathrm{ij}}, 
\end{equation}
where $A_{ij}$ and $\nu$ are the Einstein coefficient and transition frequency, respectively, with the assumption of optically thin CH$_3$OH emission. 
The total CH$_3$OH column density, $N$(CH$_3$OH), with the assumption of local-thermodynamic-equilibrium (LTE), was then determined as:
\begin{equation}
N\mathrm{(CH}_3\mathrm{OH)} = Z(T)\, \frac{N_\mathrm{u}}{g_\mathrm{u}}\,e^{\frac{E_\mathrm{u}}{k_\mathrm{B}T}}, 
\end{equation}
where $T$ is the gas temperature (assumed to be 10~K), $Z(T) = 84$ is the corresponding partition function, and $g_\mathrm{u}$ and $E_\mathrm{u}$ are the statistical weight and energy of the upper state, respectively.
Given a fractional abundance ($X$) and the core size ($\ell$), the gas density was estimated as:
\begin{equation}
\label{eq:rate}
\rho = \mu\,m_\mathrm{H}\,n\mathrm{(H}_2\mathrm{)} = \frac{\mu\,m_\mathrm{H}\,N\mathrm{(CH}_3\mathrm{OH)}}{X\,\ell}, 
\end{equation}
where $\mu = 2.35$ is the mean molecular weight per free particle, and $m_\mathrm{H}$ is the hydrogen mass. 
The mass assembly rate, as shown in Eq.~\ref{eq:rate}, is inversely proportional to the fractional abundance ($X$) and the core size ($\ell$). 
Adopting the $X \sim 10^{-6}$ from \citet{2015Boogert_review_ice} and the $\ell \sim 0.02$~pc from \citet{2021Sahu_ALMASOP_presstellar}, the total mass assembly rate ($\Dot{M}$) from $1.1$ to $3.1 \times 10^{-7}~M_\odot$\peryr. 
This value may be underestimated. 
First, the observed CH$_3$OH emission may not capture the mass under the assembling process. 
Second, the fractional abundance adopted here was measured from the ice compositions of low-mass protostellar cores.
As a result, this fractional abundance may be overestimated due to factors such as a lower reservoir of CH$_3$OH molecules within the ice and uncertainties in the desorption efficiency of mechanical removal.
A desorption efficiency of 10\% would increase the estimated mass assembly rate to $10^{-6}$~$M\odot$\peryr, aligning with the expected lifetime of a solar-like starless core of $10^6$ years \citep[e.g., ][]{2000Klessen_model_turbulence}. 
A more detailed simulation incorporating factors such as collision geometry, complex organic chemistry networks, and molecule desorption efficiency would help refine this estimate. 

\section{Core Evolution}
\label{sec:ORION2}

To test this hypothesis of turbulence-induced mass assembly, we produced the synthesized images for a starless core extracted from a filamentary dark cloud formation simulation. 
This simulation used numerical method for the study of magnetic properties of clumps/cores in the turbulence filamentary dark clouds \citep{2015Li_ORION2,2019Li_ORION2}. 
The ideal MHD simulation was performed using the multiphysics Godunov code ORION2 with block-structured Adaptive Mesh Refinement (AMR) framework \citep{2012Li_ORION2,2021Li_ORION2}. 

The simulation started with a uniform density and magnetic field at a thermal Mach number of 10 and an Alfv$\acute{e}$nic Mach number of 1. 
The initial turbulence driving over two crossing times of were applied at the largest scales with wave numbers $k = 1-2$ using the recipe described in \citet{1999Maclow_driving}. 
After the initial turbulence driving, the system was in statistical equilibrium state. 
Gravity was turned on and the system was continued to be driven to maintain a global turbulence at a thermal Mach number of 10. The system was scaled by the virial parameter of 1 when gravity was on.  
Using line width-size relation from observations \citep{2007McKee_linewidthsize}, the velocity dispersion of a Mach 10 turbulence system gives the length of the simulation region to be 4.55~pc (see Equation 1 in \citet{2019Li_ORION2}.  
The virial parameter of 1 will define the scale of mass in the system to be $3110 {\rm M_{\odot}}$ (see equation of 5 in \citet{2019Li_ORION2}).  
The Alfvenic Mach number of 1 will define the mean magnetic field strength of $31.6~\mu {\rm G}$, when virial parameter is 1 (see equation 8 in \citet{2019Li_ORION2}).
Periodic boundary conditions and isothermal equation of state were used for the simulation on a base grid of $512^3$ with 2 levels of refinement. The highest resolution is about 0.0022 pc (450~au).
The density and velocity profiles of the simulated core were extracted when its peak number density reached $1\times10^7$~cm$^{-3}$, the value for G205M3 estimated by \citet{2023Sahu_ALMASOP_prestellar}. 

We conducted radiative transfer calculations using ``SPARX (Simulation Package for Astronomical Radiative Xfer)'', which is a radiative transfer code for submillimeter/millimeter wavelengths. 
For N$_2$H$^+$, we assumed it resides in gas with number densities exceeding its critical density ($6\times10^4$~cm$^{-3}$) and adopted a fractional abundance of $10^{-9}$. 
For CH$_3$OH, we assumed it is present in regions experiencing a velocity jump ($\Delta v$) greater than 0.4 km/s, with an fractional abundance of $10^{-6}$ \citep{2015Boogert_review_ice}. 
The velocity jump ($\Delta v$) is defined by the three-dimensional velocity differences at each location. 
The dust temperature and the gas temperature were fixed to 10~K. 
The gas-to-dust mass ratio was set to 100. 

\begin{figure}[htb!]
\centering
\includegraphics[width=.9\linewidth]{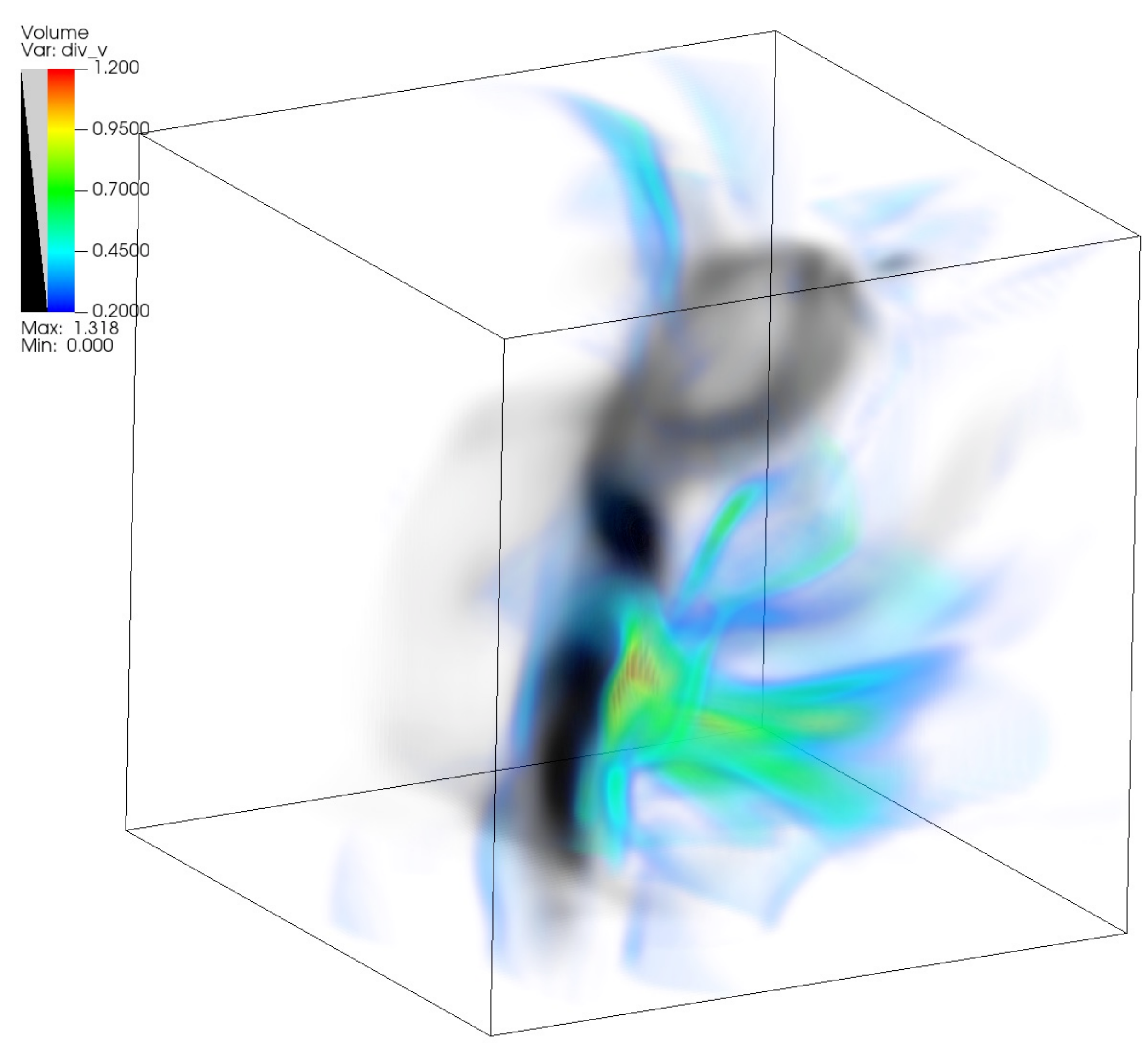}
\caption{
\label{fig:raster_3d}
Starless core evolution simulation result. 
The gray rasters illustrate the regions having gas number density greater than the critical density of N$_2$H$^+$ ($6\times10^4$~cm$^{-3}$). 
The rainbow rasters illustrate the shocked (velocity jump greater than 0.2 km/s, the sound speed) regions. 
The definition of the the velocity jump is in Section \ref{sec:methods}. 
In our simulation, gas-phase CH$_3$OH is residing in the regions having velocity jump greater than 0.4 km s$^{-1}$. 
The size of the box is approximately 0.16 pc. 
}
\end{figure}

\begin{figure*}[htb!]
\centering
\includegraphics[width=\textwidth]{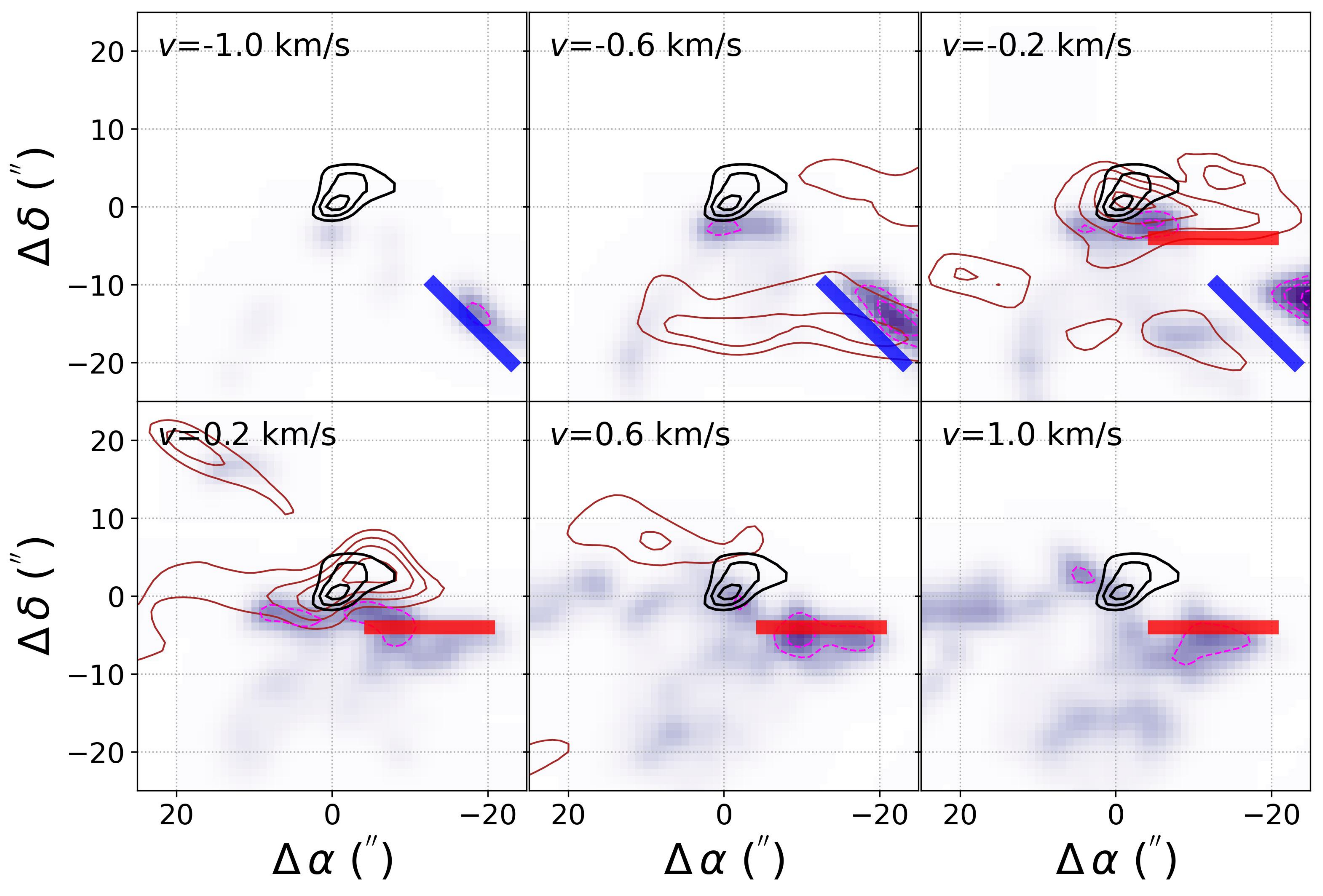}
\caption{
\label{fig:chn_shock_sim} 
Flux density channel map of the simulated core. 
The purple raster and magenta contours represents CH$_3$OH emission, the brown contours show N$_2$H$^+$ emission, and the black contours indicate the 3~mm continuum.
The contour levels for N$_2$H$^+$ (brown), CH$_3$OH (purple), and 3~mm continuum (black) are set as  [30\%, 50\%, 70\%, 90\%], [50\%, 70\%, 90\%], and [50\%, 70\%, 90\%] of the peak flux density. 
The two segments (blue and red) illustrate the interfaces between CH$_3$OH and N$_2$H$^+$ gas components having inconsistent velocity ranges. 
}
\end{figure*}

The simulation results revealed that shocks are widespread around the dense filament because of supersonic turbulence (Fig. \ref{fig:raster_3d}). 
As a demonstration of the core in the Orion cloud, the velocity jump reaches 0.4 -- 0.8 km~s$^{-1}$ (Fig. \ref{fig:chn_shock_sim}). 
Further synthetic images made under the assumption that N$_2$H$^+$ traces dense gas and CH$_3$OH traces shocked regions, successfully reproduced the observed morphologically anti-correlated and  kinematically mismatched CH$_3$OH and N$_2$H$^+$ emission.
Additionally, the CH$_3$OH emission is widespread around the N$_2$H$^+$ emission, consistent with the observed spatial distribution (Fig. \ref{fig:chn_shock_sim}).

\section{Discussion and Conclusion}
\label{sec:Disc}

Our findings reveal that, as a tracer of turbulence-induced shocks, CH$_3$OH emission captures the ongoing mass assembly from the surrounding diffuse medium onto the dense core.
This could explain why CH$_3$OH emission in starless cores often peaks at positions offset from N$_2$H$^+$ or the continuum, as observed in this study and in L1544 \citep{2014Bizzocchi_L1544_CH2DOH,2017Spezzano_L1544_ChemStructure}.
In cases such as G209S1 from our study, we observe spatially overlapping CH$_3$OH and N$_2$H$^+$ emissions.
Based on our scenario, this overlap may indicate that the shocked regions are superimposed along the line of sight, either in front of or behind the N$_2$H$^+$ gas component, or that shocks (turbulence-induced or inflow-induced) exist within the dense core itself.
Additionally, in G208N2, the observed CH$_3$OH emission could potentially be associated with nearby protostars (e.g., MMS 3, [$\Delta \alpha_{2000}$, $\Delta \delta_{2000}]\sim$[$-20\arcs$, $0\arcs$]), possibly due to outflows originating from these objects.

Given that turbulence activity in the Orion Molecular Cloud is more intense than that in other nearby star-forming clouds, such as the Taurus and Perseus clouds \cite[e.g., ][]{2022Ha_turbulence_clouds}), we examine whether there are compelling evidence indicating the association between gas-phase CH$_3$OH and turbulence activity across different clouds. 
In the starless core TMC1-CP, a ``clump-clump collision shock'' has been proposed to explain chemical variations at the cyanopolyyne peak and is discussed as a potential mechanism for CH$_3$OH production \citep{2001Dickens_TMC1,2015Soma_TMC1}.
Additionally, the line profile of CH$_3$OH in starless cores within the Taurus clouds suggests a combination of bulk motions (e.g., gradients or flows) and supersonic turbulence \citep{2020Scibelli_ColdCoresTaurus_COMs}.
Moreover, surveys of starless cores in the Taurus \citep{2020Scibelli_ColdCoresTaurus_COMs} and Perseus \citep{2024Scibelli_Perseus_COM} clouds indicate low excitation temperatures for CH$_3$OH ($\lesssim$ 10 K), supporting that CH$_3$OH is desorbed via mechanical removal.
Further survey-type observations with high spatial and spectral resolution are essential to determine whether turbulence-induced mass assembly is also at work in these clouds, as observed in the Orion cloud, especially given the varying properties of star-forming regions (e.g., high- vs. low-mass regions and differences in dust temperature).

A further question is the timescale over which CH$_3$OH gas can persist in the gas phase before, for instance, depleting back onto dust grains. 
A short timescale would suggest that the observed CH$_3$OH emission represents a snapshot of an active turbulence-induced mass assembly process. 
Conversely, a longer survival timescale would imply the presence of an extended CH$_3$OH gas component, reflecting the widespread turbulence activity in starless cores (e.g., Fig. \ref{fig:raster_3d}). 
Future simulations that incorporate chemical networks, alongside laboratory experiments, could offer critical insights into this phenomenon. 
Moreover, comparisons between single-dish and interferometric observations may help identify and characterize such extended CH$_3$OH components.

\newpage
\clearpage



\acknowledgments
This paper makes use of the following ALMA data: ADS/JAO.ALMA\#\#2023.1.01067.S. ALMA is a partnership of ESO (representing its member states), NSF (USA), and NINS (Japan), together with NRC (Canada), MOST and ASIAA (Taiwan), and KASI (Republic of Korea), in cooperation with the Republic of Chile. The Joint ALMA Observatory is operated by ESO, AUI/NRAO, and NAOJ.
Tie Liu acknowledges the supports by the National Key R\&D Program of China (No. 2022YFA1603101), National Natural Science Foundation of China (NSFC) through grants No.12073061 and No.12122307. 
S.-Y. H. and S.-Y.L. acknowledge grants from the National Science and Technology Council of Taiwan (112-2112-M-001-060- and 113-2112-M-001-004-).
P.S. Li acknowledges the supports by the National Key R\&D Program of China (No. 2022YFA1603101) and NSFC through grant No. 1241101426. 
D.S. acknowledges the support from Ramanujan Fellowship (SERB, RJF/2021/000116) and PRL. 
X.L. acknowledges the support of the Strategic Priority Research Program of the Chinese Academy of Sciences  under Grant No. XDB0800303.

\software{
astropy \citep{astropy:2013, astropy:2018},
\texttt{CASA} \citep{casa:2007,casa:2022},
\texttt{CARTA}  \citep{2021Comrie_CARTA}, 
\texttt{SPARX}
}

\appendix

\section{Supplementary figures}
\resetapptablenumbers

\begin{figure*}[h]
\centering
\includegraphics[width=.99\textwidth]{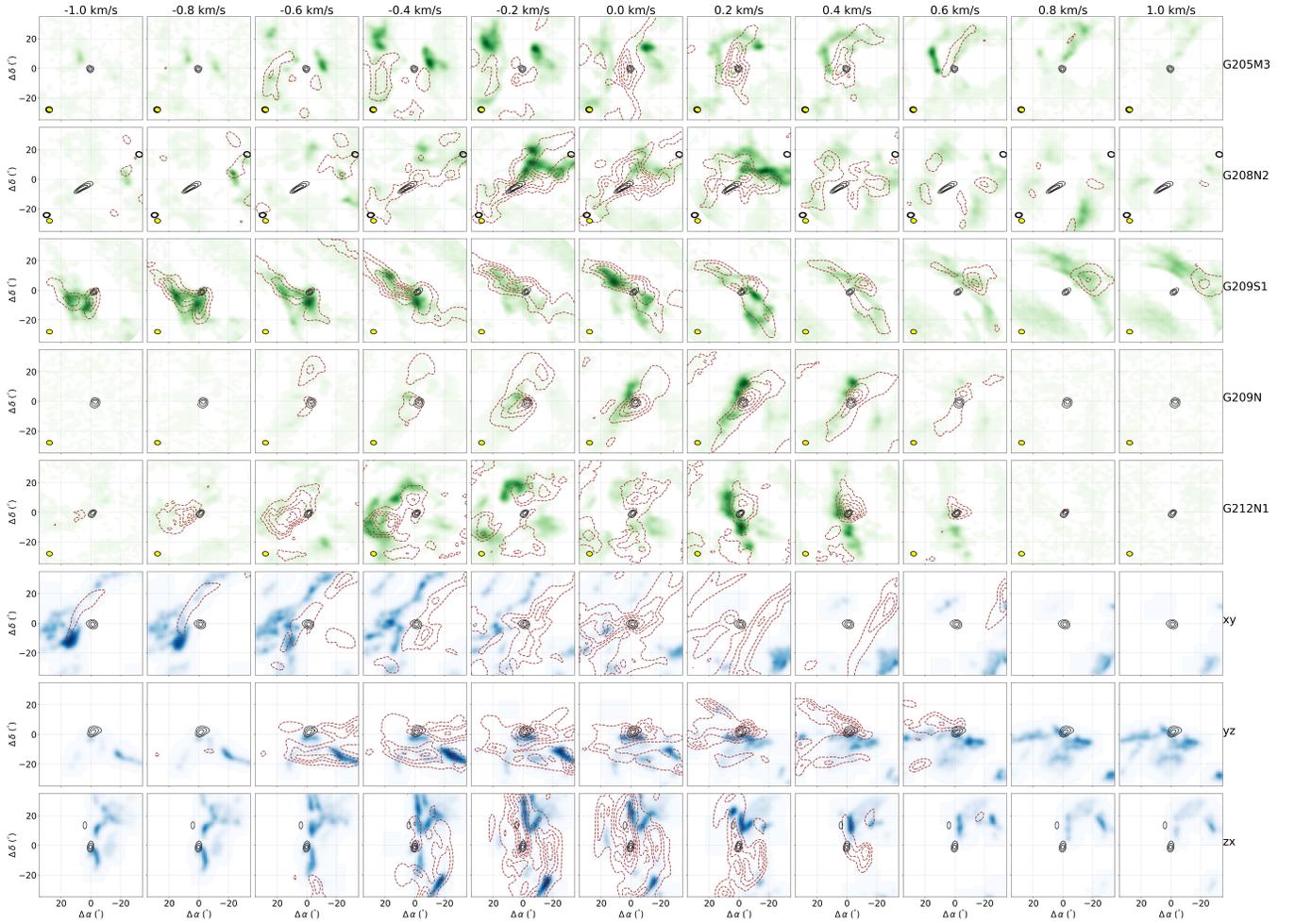}
\caption{
\label{fig:chn_map}
Channel maps of the observations and simulations.
The top five rows display channel maps of the five observed fields, with source names labeled on the right of each row.
The bottom three rows show channel maps of a simulated core from three different projected lines of sight.
The CH$_3$OH emission is represented in green and blue raster colors, N$_2$H$^+$ emission is shown as brown contours, and the 3~mm continuum is indicated by black contours.
Contour levels for N$_2$H$^+$ and the continuum are set at [10\%, 30\%, 50\%, 70\%, 90\%] and [60\%, 75\%, 90\%] of their respective peak flux densities for each row.
}
\end{figure*}

\begin{figure}[htb!]
\centering
\includegraphics[width=.9\linewidth]{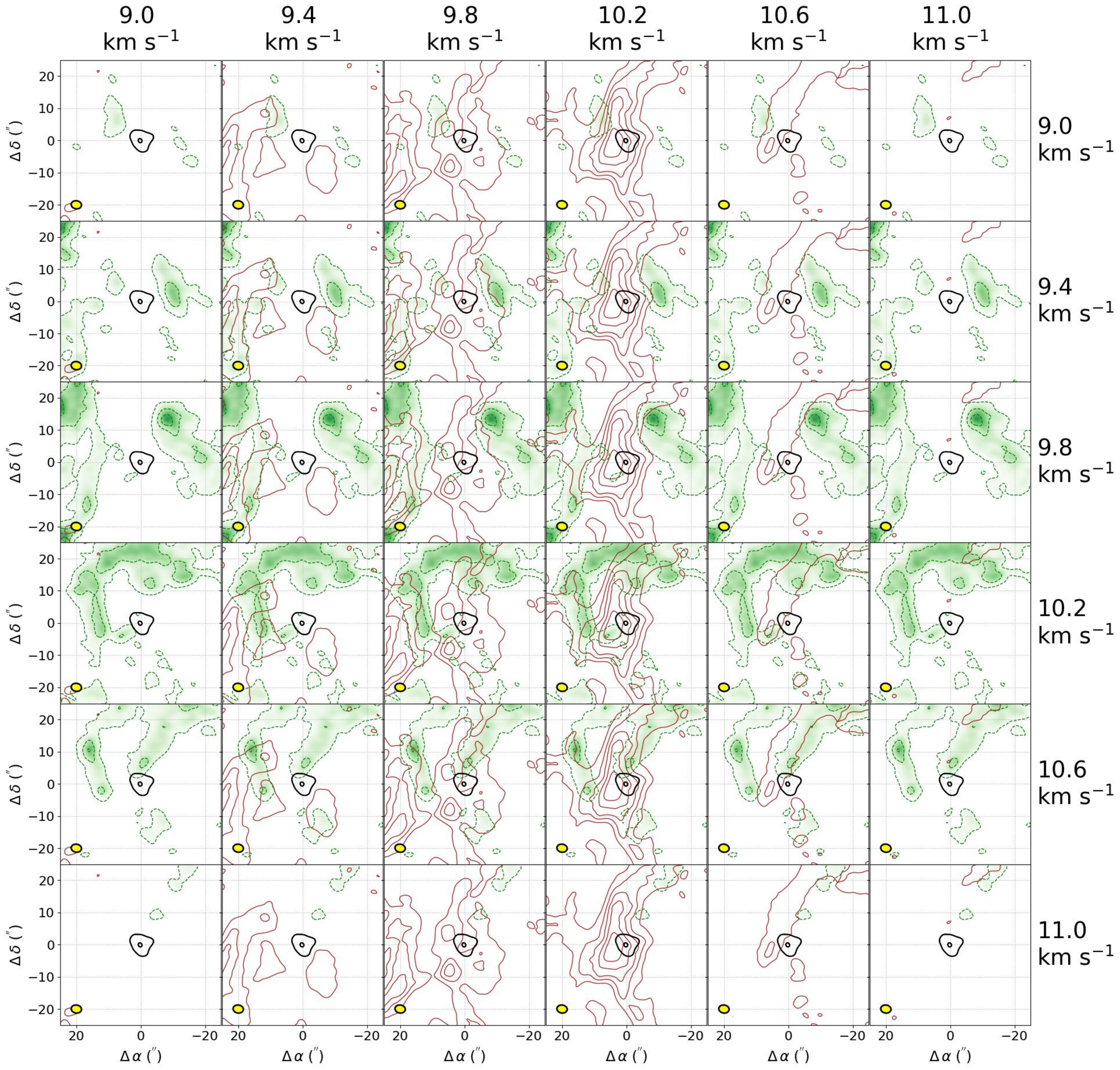}
\caption{
\label{fig:matrix_G205M3}
Channel map matrix of G205.46-14.56M3. 
The green raster represents CH$_3$OH emission, the brown contours show N$_2$H$^+$ emission, and the black contours indicate the 3~mm continuum.
The contour levels for N$_2$H$^+$ (brown), CH$_3$OH (green), and 3~mm continuum (black) are set as  [5, 35, 65, 95, 125] $\sigma$, [5, 15, 25, 35, 45] $\sigma$, and [5, 15, 25, 35] $\sigma$. 
The origins of the coordinates are set at the 3~mm continuum peak position. 
The velocities labeled on the top and the right are for N$_2$H$^+$ and CH$_3$OH, respectively. 
}
\end{figure}

\begin{figure*}[htb!]
\centering
\includegraphics[width=\textwidth]{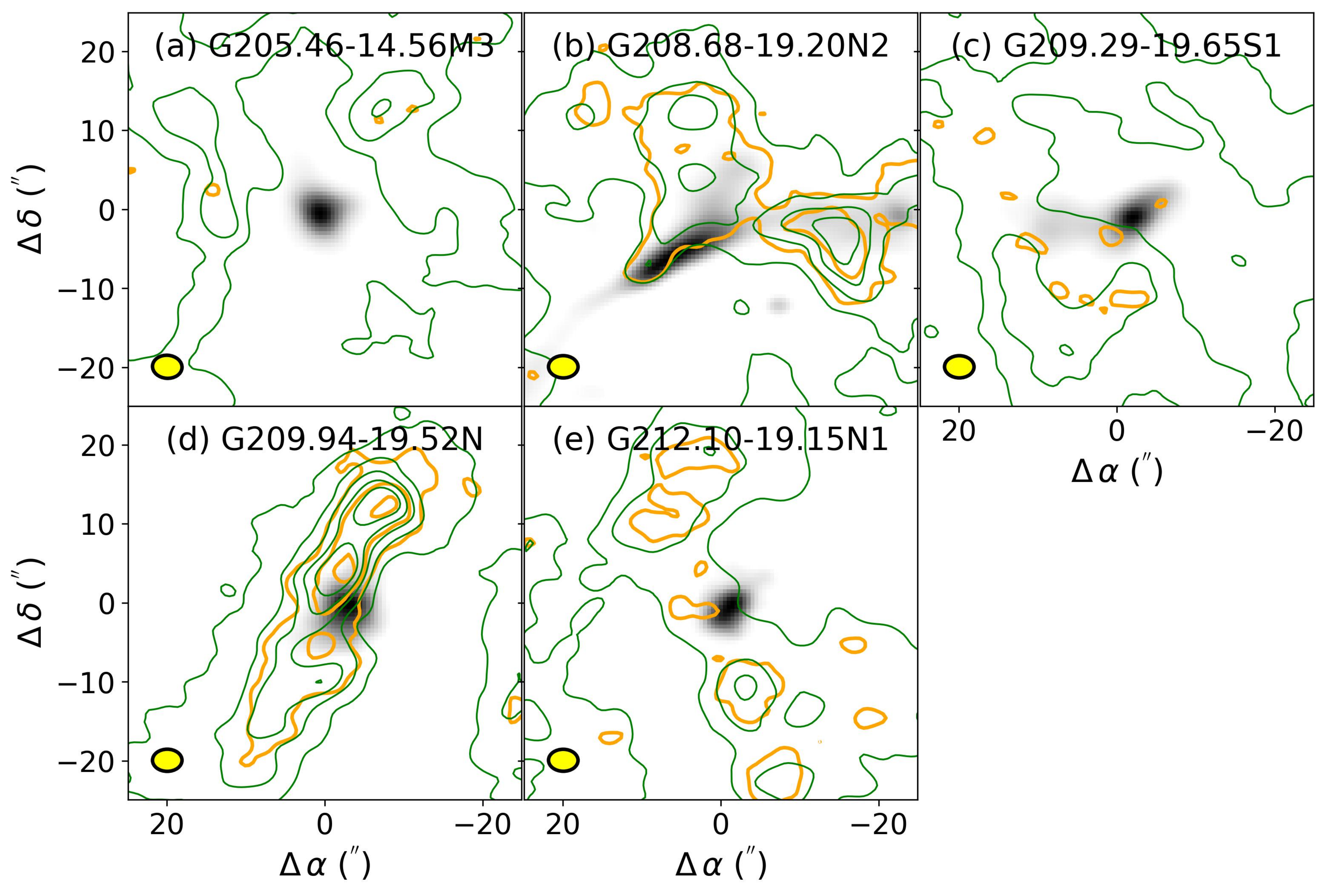}
\caption{
\label{fig:mom0_CH3OH_20K}
Integrated intensities of the CH$_3$OH $E_\mathrm{u}=7$~K transition ($2_{0,2} – 1_{0,1}~A^+$, shown as green contours), the CH$_3$OH $E_\mathrm{u}=20$~K transition ($2_{0,2} – 1_{0,1}~E_1$, shown as orange contours), and the 3~mm continuum (black rasters).
The large dashed circles indicate the field of view, while the small ellipses in the bottom left corner represent the synthesized beam.
The contour levels for the CH$_3$OH 7K transition (green) are set at [5, 25, 45, 65, 85] $\sigma$.
The contour levels for the CH$_3$OH 20K transition (orange) are set at [5, 15, 25] $\sigma$.
}
\end{figure*}

Fig. \ref{fig:chn_map} shows the channel maps of the observations and simulations for completeness of the report. 
Fig. \ref{fig:matrix_G205M3} shows the channel map matrix of G205.46-14.56M3, which we used for visually inspecting the shock interface. 
Fig. \ref{fig:mom0_CH3OH_20K} shows the integrated intensities of the CH$_3$OH $E_\mathrm{u}=7$~K transition and  $E_\mathrm{u}=20$~K transitions and the 3~mm continuum.

\bibliographystyle{aasjournal}

\begin{thebibliography}{}
\expandafter\ifx\csname natexlab\endcsname\relax\def\natexlab#1{#1}\fi
\providecommand{\url}[1]{\href{#1}{#1}}
\providecommand{\dodoi}[1]{doi:~\href{http://doi.org/#1}{\nolinkurl{#1}}}
\providecommand{\doeprint}[1]{\href{http://ascl.net/#1}{\nolinkurl{http://ascl.net/#1}}}
\providecommand{\doarXiv}[1]{\href{https://arxiv.org/abs/#1}{\nolinkurl{https://arxiv.org/abs/#1}}}

\bibitem[{{Astropy Collaboration} {et~al.}(2013){Astropy Collaboration}, {Robitaille}, {Tollerud}, {Greenfield}, {Droettboom}, {Bray}, {Aldcroft}, {Davis}, {Ginsburg}, {Price-Whelan}, {Kerzendorf}, {Conley}, {Crighton}, {Barbary}, {Muna}, {Ferguson}, {Grollier}, {Parikh}, {Nair}, {Unther}, {Deil}, {Woillez}, {Conseil}, {Kramer}, {Turner}, {Singer}, {Fox}, {Weaver}, {Zabalza}, {Edwards}, {Azalee Bostroem}, {Burke}, {Casey}, {Crawford}, {Dencheva}, {Ely}, {Jenness}, {Labrie}, {Lim}, {Pierfederici}, {Pontzen}, {Ptak}, {Refsdal}, {Servillat}, \& {Streicher}}]{astropy:2013}
{Astropy Collaboration}, {Robitaille}, T.~P., {Tollerud}, E.~J., {et~al.} 2013, \aap, 558, A33, \dodoi{10.1051/0004-6361/201322068}

\bibitem[{Balsara {et~al.}(2001)Balsara, Ward-Thompson, \& Crutcher}]{2001Balsara_turbulent-MHD-model}
Balsara, D., Ward-Thompson, D., \& Crutcher, R.~M. 2001, Monthly Notices of the Royal Astronomical Society, 327, 715, \dodoi{10.1046/j.1365-8711.2001.04787.x}

\bibitem[{Bergin \& Tafalla(2007)}]{2007Bergin_review}
Bergin, E.~A., \& Tafalla, M. 2007, Annual Review of Astronomy and Astrophysics, 45, 339, \dodoi{10.1146/annurev.astro.45.071206.100404}

\bibitem[{Bizzocchi {et~al.}(2014)Bizzocchi, Caselli, Spezzano, \& Leonardo}]{2014Bizzocchi_L1544_CH2DOH}
Bizzocchi, L., Caselli, P., Spezzano, S., \& Leonardo, E. 2014, Astronomy and Astrophysics, 569, A27, \dodoi{10.1051/0004-6361/201423858}

\bibitem[{Boogert {et~al.}(2015)Boogert, Gerakines, \& Whittet}]{2015Boogert_review_ice}
Boogert, A. C.~A., Gerakines, P.~A., \& Whittet, D. C.~B. 2015, Annual Review of Astronomy and Astrophysics, Vol 53, 53, 541, \dodoi{10.1146/annurev-astro-082214-122348}

\bibitem[{{CASA Team} {et~al.}(2022){CASA Team}, Bean, Bhatnagar, Castro, Donovan~Meyer, Emonts, Garcia, Garwood, Golap, Gonzalez~Villalba, Harris, Hayashi, Hoskins, Hsieh, Jagannathan, Kawasaki, Keimpema, Kettenis, Lopez, Marvil, Masters, McNichols, Mehringer, Miel, Moellenbrock, Montesino, Nakazato, Ott, Petry, Pokorny, Raba, Rau, Schiebel, Schweighart, Sekhar, Shimada, Small, Steeb, Sugimoto, Suoranta, Tsutsumi, van Bemmel, Verkouter, Wells, Xiong, Szomoru, Griffith, Glendenning, \& Kern}]{casa:2022}
{CASA Team}, Bean, B., Bhatnagar, S., {et~al.} 2022, Publications of the Astronomical Society of the Pacific, 134, 114501, \dodoi{10.1088/1538-3873/ac9642}

\bibitem[{Chuang {et~al.}(2018)Chuang, Fedoseev, Qasim, Ioppolo, van Dishoeck, \& Linnartz}]{2018Chuang_reactive-desorption}
Chuang, K.~J., Fedoseev, G., Qasim, D., {et~al.} 2018, The Astrophysical Journal, 853, 102, \dodoi{10.3847/1538-4357/aaa24e}

\bibitem[{Comrie {et~al.}(2021)Comrie, Wang, Hsu, Moraghan, Harris, Pang, Pińska, Chiang, Simmonds, Chang, Jan, \& Lin}]{2021Comrie_CARTA}
Comrie, A., Wang, K.-S., Hsu, S.-C., {et~al.} 2021, Astrophysics Source Code Library, ascl:2103.031.
\newblock \url{https://ui.adsabs.harvard.edu/abs/2021ascl.soft03031C}

\bibitem[{Dartois {et~al.}(2020)Dartois, Chabot, Bacmann, Boduch, Domaracka, \& Rothard}]{2019Dartois_CR_sputtering}
Dartois, E., Chabot, M., Bacmann, A., {et~al.} 2020, Astronomy and Astrophysics, 634, A103, \dodoi{10.1051/0004-6361/201936934}

\bibitem[{Dickens {et~al.}(2001)Dickens, Langer, \& Velusamy}]{2001Dickens_TMC1}
Dickens, J.~E., Langer, W.~D., \& Velusamy, T. 2001, The Astrophysical Journal, 558, 693, \dodoi{10.1086/322292}

\bibitem[{Dutta {et~al.}(2020)Dutta, Lee, Liu, Hirano, Liu, Tatematsu, Kim, Shang, Sahu, Kim, Moraghan, Jhan, Hsu, Evans, Johnstone, Ward-Thompson, Kuan, Lee, Lee, Traficante, Juvela, Vastel, Zhang, Sanhueza, Soam, Kwon, Bronfman, Eden, Goldsmith, He, Wu, Pelkonen, Qin, Li, \& Li}]{2020Dutta_ALMASOP}
Dutta, S., Lee, C.-F., Liu, T., {et~al.} 2020, \apjs, 251, 20, \dodoi{10.3847/1538-4365/abba26}

\bibitem[{Fielder {et~al.}(2024)Fielder, Kirk, Dunham, \& Offner}]{2024Fielder_G205M3}
Fielder, S.~D., Kirk, H., Dunham, M.~M., \& Offner, S. S.~R. 2024, The Astrophysical Journal, 968, 10, \dodoi{10.3847/1538-4357/ad3d56}

\bibitem[{Garrod {et~al.}(2007)Garrod, Wakelam, \& Herbst}]{2007Garrod_reactive-desorption}
Garrod, R.~T., Wakelam, V., \& Herbst, E. 2007, Astronomy and Astrophysics, 467, 1103, \dodoi{10.1051/0004-6361:20066704}

\bibitem[{Ha {et~al.}(2022)Ha, Li, Kounkel, Xu, Li, \& Zheng}]{2022Ha_turbulence_clouds}
Ha, T., Li, Y., Kounkel, M., {et~al.} 2022, The Astrophysical Journal, 934, 7, \dodoi{10.3847/1538-4357/ac76bf}

\bibitem[{Hartmann {et~al.}(2001)Hartmann, Ballesteros-Paredes, \& Bergin}]{2001_Hartmann}
Hartmann, L., Ballesteros-Paredes, J., \& Bergin, E.~A. 2001, The Astrophysical Journal, 562, 852, \dodoi{10.1086/323863}

\bibitem[{Herbst \& van Dishoeck(2009)}]{2009Herbst_COM_review}
Herbst, E., \& van Dishoeck, E.~F. 2009, Annual Review of Astronomy and Astrophysics, Vol 47, 47, 427, \dodoi{10.1146/annurev-astro-082708-101654}

\bibitem[{Hirano {et~al.}(2024)Hirano, Sahu, Liu, Liu, Tatematsu, Dutta, Li, Lee, Li, Hsu, Lin, Johnstone, Bronfman, Chen, Eden, Kuan, Kwon, Lee, Liu, Rawlings, Ristorcelli, \& Traficante}]{2024Hirano_G208.68-19.20N2_dense}
Hirano, N., Sahu, D., Liu, S.-Y., {et~al.} 2024, The Astrophysical Journal, 961, 123, \dodoi{10.3847/1538-4357/ad09e2}

\bibitem[{Hsu {et~al.}(2023)Hsu, Liu, Johnstone, Liu, Bronfman, Chen, Dutta, Eden, Evans, Hirano, Juvela, Kuan, Kwon, Lee, Lee, Lee, Li, Liu, Liu, Luo, Qin, Rawlings, Sahu, Sanhueza, Shang, Tatematsu, \& Yang}]{2023Hsu_ALMASOP}
Hsu, S.-Y., Liu, S.-Y., Johnstone, D., {et~al.} 2023, The Astrophysical Journal, 956, 120, \dodoi{10.3847/1538-4357/acefcf}

\bibitem[{Kalvāns \& Silsbee(2022)}]{2022Kalvans_desorption}
Kalvāns, J., \& Silsbee, K. 2022, Monthly Notices of the Royal Astronomical Society, 515, 785, \dodoi{10.1093/mnras/stac1792}

\bibitem[{Keown {et~al.}(2017)Keown, Di~Francesco, Kirk, Friesen, Pineda, Rosolowsky, Ginsburg, Offner, Caselli, Alves, Chacón-Tanarro, Punanova, Redaelli, Seo, Matzner, Chun-Yuan~Chen, Goodman, Chen, Shirley, Singh, Arce, Martin, \& Myers}]{2017Keown_CepheusL1251}
Keown, J., Di~Francesco, J., Kirk, H., {et~al.} 2017, The Astrophysical Journal, 850, 3, \dodoi{10.3847/1538-4357/aa93ec}

\bibitem[{Klessen {et~al.}(2000)Klessen, Heitsch, \& Mac~Low}]{2000Klessen_model_turbulence}
Klessen, R.~S., Heitsch, F., \& Mac~Low, M.-M. 2000, The Astrophysical Journal, 535, 887, \dodoi{10.1086/308891}

\bibitem[{Larson(1981)}]{1981Larson_turbulence}
Larson, R.~B. 1981, Monthly Notices of the Royal Astronomical Society, 194, 809, \dodoi{10.1093/mnras/194.4.809}

\bibitem[{Lazarian \& Cho(2004)}]{2004Lazarian_review}
Lazarian, A., \& Cho, J. 2004, Astrophysics and Space Science, 292, 29, \dodoi{10.1023/B:ASTR.0000044998.74603.91}

\bibitem[{Li {et~al.}(2021)Li, Cunningham, Gaches, Klein, Krumholz, Lee, McKee, Offner, Rosen, \& Skinner}]{2021Li_ORION2}
Li, P., Cunningham, A., Gaches, B., {et~al.} 2021, The Journal of Open Source Software, 6, 3771, \dodoi{10.21105/joss.03771}

\bibitem[{Li \& Klein(2019)}]{2019Li_ORION2}
Li, P.~S., \& Klein, R.~I. 2019, Monthly Notices of the Royal Astronomical Society, 485, 4509, \dodoi{10.1093/mnras/stz653}

\bibitem[{Li {et~al.}(2012)Li, Martin, Klein, \& McKee}]{2012Li_ORION2}
Li, P.~S., Martin, D.~F., Klein, R.~I., \& McKee, C.~F. 2012, The Astrophysical Journal, 745, 139, \dodoi{10.1088/0004-637X/745/2/139}

\bibitem[{Li {et~al.}(2015)Li, McKee, \& Klein}]{2015Li_ORION2}
Li, P.~S., McKee, C.~F., \& Klein, R.~I. 2015, Monthly Notices of the Royal Astronomical Society, 452, 2500, \dodoi{10.1093/mnras/stv1437}

\bibitem[{{Mac Low}(1999)}]{1999Maclow_driving}
{Mac Low}, M.-M. 1999, \apj, 524, 169, \dodoi{10.1086/307784}

\bibitem[{{McKee} \& {Ostriker}(2007)}]{2007McKee_linewidthsize}
{McKee}, C.~F., \& {Ostriker}, E.~C. 2007, \araa, 45, 565, \dodoi{10.1146/annurev.astro.45.051806.110602}

\bibitem[{{McMullin} {et~al.}(2007){McMullin}, Waters, Schiebel, Young, \& Golap}]{casa:2007}
{McMullin}, J.~P., Waters, B., Schiebel, D., Young, W., \& Golap, K. 2007, in Astronomical Data Analysis Software and Systems XVI, ed. R.~A. Shaw, F.~Hill, \& D.~J. Bell, Vol. 376, San Francisco, Astronomical Society of the Pacific (ASP Conference Series)

\bibitem[{Nakai {et~al.}(2023)Nakai, Sameera, Furuya, Hidaka, Ishibashi, \& Watanabe}]{2023Nakai_CH3OH_formation}
Nakai, Y., Sameera, W. M.~C., Furuya, K., {et~al.} 2023, The Astrophysical Journal, 953, 162, \dodoi{10.3847/1538-4357/accf95}

\bibitem[{Ohashi {et~al.}(2016)Ohashi, Tatematsu, Sanhueza, Luong, Hirota, Choi, \& Mizuno}]{2016Ohashi_TUKH122_turbulence}
Ohashi, S., Tatematsu, K., Sanhueza, P., {et~al.} 2016, Monthly Notices of the Royal Astronomical Society, 459, 4130, \dodoi{10.1093/mnras/stw856}

\bibitem[{Padoan \& Nordlund(1999)}]{1999_Padoan}
Padoan, P., \& Nordlund, A. 1999, The Astrophysical Journal, 526, 279, \dodoi{10.1086/307956}

\bibitem[{{Price-Whelan} {et~al.}(2018){Price-Whelan}, {Sip{\H{o}}cz}, {G{\"u}nther}, {Lim}, {Crawford}, {Conseil}, {Shupe}, {Craig}, {Dencheva}, {Ginsburg}, {VanderPlas}, {Bradley}, {P{\'e}rez-Su{\'a}rez}, {de Val-Borro}, {Paper Contributors}, {Aldcroft}, {Cruz}, {Robitaille}, {Tollerud}, {Coordination Committee}, {Ardelean}, {Babej}, {Bach}, {Bachetti}, {Bakanov}, {Bamford}, {Barentsen}, {Barmby}, {Baumbach}, {Berry}, {Biscani}, {Boquien}, {Bostroem}, {Bouma}, {Brammer}, {Bray}, {Breytenbach}, {Buddelmeijer}, {Burke}, {Calderone}, {Cano Rodr{\'\i}guez}, {Cara}, {Cardoso}, {Cheedella}, {Copin}, {Corrales}, {Crichton}, {D{\textquoteright}Avella}, {Deil}, {Depagne}, {Dietrich}, {Donath}, {Droettboom}, {Earl}, {Erben}, {Fabbro}, {Ferreira}, {Finethy}, {Fox}, {Garrison}, {Gibbons}, {Goldstein}, {Gommers}, {Greco}, {Greenfield}, {Groener}, {Grollier}, {Hagen}, {Hirst}, {Homeier}, {Horton}, {Hosseinzadeh}, {Hu}, {Hunkeler}, {Ivezi{\'c}}, {Jain}, {Jenness}, {Kanarek}, {Kendrew}, {Kern}, {Kerzendorf}, {Khvalko},
  {King}, {Kirkby}, {Kulkarni}, {Kumar}, {Lee}, {Lenz}, {Littlefair}, {Ma}, {Macleod}, {Mastropietro}, {McCully}, {Montagnac}, {Morris}, {Mueller}, {Mumford}, {Muna}, {Murphy}, {Nelson}, {Nguyen}, {Ninan}, {N{\"o}the}, {Ogaz}, {Oh}, {Parejko}, {Parley}, {Pascual}, {Patil}, {Patil}, {Plunkett}, {Prochaska}, {Rastogi}, {Reddy Janga}, {Sabater}, {Sakurikar}, {Seifert}, {Sherbert}, {Sherwood-Taylor}, {Shih}, {Sick}, {Silbiger}, {Singanamalla}, {Singer}, {Sladen}, {Sooley}, {Sornarajah}, {Streicher}, {Teuben}, {Thomas}, {Tremblay}, {Turner}, {Terr{\'o}n}, {van Kerkwijk}, {de la Vega}, {Watkins}, {Weaver}, {Whitmore}, {Woillez}, {Zabalza}, \& {Contributors}}]{astropy:2018}
{Price-Whelan}, A.~M., {Sip{\H{o}}cz}, B.~M., {G{\"u}nther}, H.~M., {et~al.} 2018, \aj, 156, 123, \dodoi{10.3847/1538-3881/aabc4f}

\bibitem[{Sahu {et~al.}(2021)Sahu, Liu, Liu, Evans, Hirano, Tatematsu, Lee, Kim, Dutta, Alina, Bronfman, Cunningham, Eden, Garay, Goldsmith, He, Hsu, Jhan, Johnstone, Juvela, Kim, Kuan, Kwon, Lee, Lee, Li, Li, Li, Luo, Montillaud, Moraghan, Pelkonen, Qin, Ristorcelli, Sanhueza, Shang, Shen, Soam, Wu, Zhang, \& Zhou}]{2021Sahu_ALMASOP_presstellar}
Sahu, D., Liu, S.-Y., Liu, T., {et~al.} 2021, The Astrophysical Journal, 907, L15, \dodoi{10.3847/2041-8213/abd3aa}

\bibitem[{Sahu {et~al.}(2023)Sahu, Liu, Johnstone, Liu, Evans, Hirano, Tatematsu, Di~Francesco, Lee, Kim, Dutta, Hsu, Li, Luo, Sanhueza, Shang, Traficante, Juvela, Lee, Eden, Goldsmith, Bronfman, Kwon, Lee, Kuan, \& Ristorcelli}]{2023Sahu_ALMASOP_prestellar}
Sahu, D., Liu, S.-Y., Johnstone, D., {et~al.} 2023, The Astrophysical Journal, 945, 156, \dodoi{10.3847/1538-4357/acbc26}

\bibitem[{Scibelli \& Shirley(2020)}]{2020Scibelli_ColdCoresTaurus_COMs}
Scibelli, S., \& Shirley, Y. 2020, The Astrophysical Journal, 891, 73, \dodoi{10.3847/1538-4357/ab7375}

\bibitem[{Scibelli {et~al.}(2024)Scibelli, Shirley, Megías, \& Jiménez-Serra}]{2024Scibelli_Perseus_COM}
Scibelli, S., Shirley, Y., Megías, A., \& Jiménez-Serra, I. 2024, arXiv e-prints, arXiv:2408.11613, \dodoi{10.48550/arXiv.2408.11613}

\bibitem[{Shu {et~al.}(1987)Shu, Adams, \& Lizano}]{1987_Shu}
Shu, F.~H., Adams, F.~C., \& Lizano, S. 1987, Annual Review of Astronomy and Astrophysics, 25, 23, \dodoi{10.1146/annurev.aa.25.090187.000323}

\bibitem[{Soma {et~al.}(2015)Soma, Sakai, Watanabe, \& Yamamoto}]{2015Soma_TMC1}
Soma, T., Sakai, N., Watanabe, Y., \& Yamamoto, S. 2015, The Astrophysical Journal, 802, 74, \dodoi{10.1088/0004-637x/802/2/74}

\bibitem[{Spezzano {et~al.}(2017)Spezzano, Caselli, Bizzocchi, Giuliano, \& Lattanzi}]{2017Spezzano_L1544_ChemStructure}
Spezzano, S., Caselli, P., Bizzocchi, L., Giuliano, B.~M., \& Lattanzi, V. 2017, Astronomy and Astrophysics, 606, A82, \dodoi{10.1051/0004-6361/201731262}

\bibitem[{Vasyunin {et~al.}(2017)Vasyunin, Caselli, Dulieu, \& Jiménez-Serra}]{2017Vasyunin_reactive-desorption}
Vasyunin, A.~I., Caselli, P., Dulieu, F., \& Jiménez-Serra, I. 2017, The Astrophysical Journal, 842, 33, \dodoi{10.3847/1538-4357/aa72ec}

\bibitem[{Vasyunin \& Herbst(2013)}]{2013Vasyunin_reactive-desorption}
Vasyunin, A.~I., \& Herbst, E. 2013, The Astrophysical Journal, 769, 34, \dodoi{10.1088/0004-637X/769/1/34}

\bibitem[{Wakelam {et~al.}(2021)Wakelam, Dartois, Chabot, Spezzano, Navarro-Almaida, Loison, \& Fuente}]{2021Wakelam_CR_sputtering}
Wakelam, V., Dartois, E., Chabot, M., {et~al.} 2021, Astronomy and Astrophysics, 652, A63, \dodoi{10.1051/0004-6361/202039855}

\bibitem[{Yamamoto(2016)}]{2016Yamamoto_book}
Yamamoto, S. 2016, Introduction to Astrochemistry: Chemical Evolution from Interstellar Clouds to Star and Planet Formation (Springer Japan).
\newblock \url{https://books.google.com.tw/books?id=Q0sbswEACAAJ}

\end{thebibliography}




\end{CJK*}
\end{document}